\documentclass[12pt, a4paper, dvips]{article}
\usepackage{amsmath}
\usepackage{amsfonts}
\usepackage{amssymb}
\usepackage{graphicx}

\usepackage{float}
\usepackage{theorem}

\numberwithin{equation}{section}
\numberwithin{table}{section}

\newtheorem{Th}{Theorem}

\begin{document}  

\begin{titlepage}

\vspace{1cm}
\begin{center}
\Huge \it Historical and other Remarks on Hidden Symmetries\footnote{Summer
School on hidden Symmetries and Higgs Phenomena, Zuoz (Engadin), Switzerland,
 August 16-22, 1998.}\\
\vspace{1cm}
\large 
 (Norbert Straumann, University of Z\"urich)

\end{center}

\vspace{5cm}

\begin{abstract}
Apart from a few remarks on lattice systems with global or gauge symmetries,
most of this talk is devoted to some interesting ancient examples of symmetries
and their breakdowns in elasticity theory and hydrodynamics. Since Galois
Theory is in many ways the origin of group theory as a tool to analyse (hidden)
symmetries, a brief review of this profound mathematical theory is also given.
\end{abstract}
\end{titlepage}

\subsection*{Introductory Remarks}
The organisers have asked me to entertain you in an evening lecture with some
historical episodes, related to symmetries and their spontaneous breakdowns,
the main theme of this Summer School. This is indeed a fascinating subject.
I shall begin with ancient examples, connected with great names, like Euler,
Galois, Jacobi, $\cdot \cdot \cdot$. In a second part of my talk I would, 
however, like to 
add a few non-historical remarks which are relevant for (lattice) field theory.
These should be regarded as supplements to the lectures by Lochlainn
O'Raifeartaigh, Daniel Loss, and others.

In one hour I cannot cover the various topics in any depth. To compensate
for this, I shall add a few references to sources which I find interesting
and pleasant to read.

\tableofcontents
\newpage

\section{Euler's instability analysis of rods under longitudinal compressional
forces}
Leonhard Euler, the man who created more mathematics than anybody else in 
history, was also one of the leading figures in the development of elasticity
theory \cite{Tru60}. In one of his later works on this subject, "Determinatio
onerum, quae columnae gestare valent" (Determination of loads which may be 
supported by columns), submitted to the Academy in Petersburg in 1776, Euler
studies again the following problem.\\

Consider a thin (metal) rod of length $L$ and circular cross section of radius
$R$. Assume that the rod is clamped at both ends and subjected to a
compressional force $F$ directed along the rod axis ($z$ axis). We denote 
the deflections of the rod in the transversal $x$ and $y$ directions as 
functions of $z$ by $X(z)$ and $Y(z)$, respectively. For {\em small} 
deflections Euler derives from the theory of elasticity the following 
differential equations
\begin{eqnarray}
     IE~\frac{d^{4} X}{dz^{4}}+F~\frac{d^{2} X}{dz^{2}} &=& 0~,\nonumber \\
     IE~\frac{d^{4} Y}{dz^{4}}+F~\frac{d^{2} Y}{dz^{2}} &=& 0~,
     \label{eq:deflection}
\end{eqnarray}
where $E$ is the Young's modulus (which was actually introduced already by 
Euler in the paper mentioned above) and $I$ is the moment of inertia, 
$I=1/4~ \pi R^{4}$. (For a textbook
derivation of these equations, see \cite{Lan}.)\\

The boundary conditions of the clamped rod are
\begin{eqnarray}
     && X(0)=X(L)=0~, \nonumber \\
     && \frac{dX}{dz}(0)=\frac{dX}{dz}(L)=0~,\label{eq:bcond}
\end{eqnarray}
and similarly for $Y$.\\

\begin{figure}[h]
\begin{center}
\includegraphics[height=0.25\textheight, angle=0]{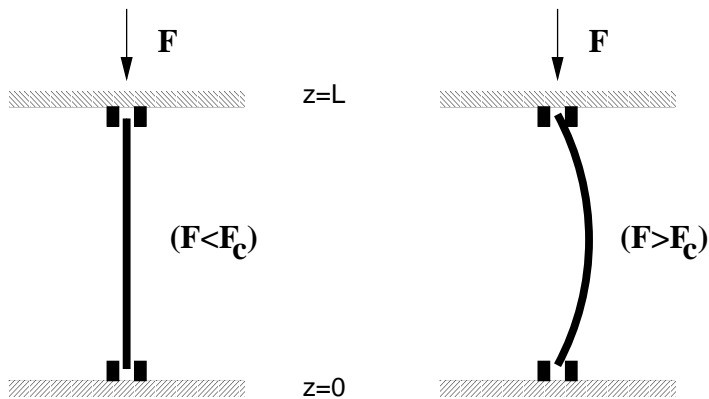}
\caption{\small{Elastic instability of a longitudinally compressed rod.}}
\end{center}
\label{fig:rod}
\end{figure}

Clearly, as long as the force $F$ is sufficiently small, the rod will be 
straight; that is, the only solution of (\ref{eq:deflection}) and 
(\ref{eq:bcond}) will be $X(z)=Y(z)\equiv 0$, and the rod is stable. However,
if $F$ is increased there will be a critical value $F_{c}$, above which the 
rod is unstable against small perturbations from straightness and will bend
(see Fig.\ref{fig:rod}).
Although in general the deflection will be large, equations 
(\ref{eq:deflection}) can still be used to find the critical value $F_{c}$.
We just have to find out when (\ref{eq:deflection}) will have a nontrivial
solution $X(z)$, satisfying the boundary conditions (\ref{eq:bcond}).

The most general solution of (\ref{eq:deflection}) for $X(z)$ is
\begin{equation}
     X(z)=A+Bz+C \sin (kz)+D \cos (kz)~,~~k=\sqrt{\frac{F}{EI}}~.
     \label{eq:gensol}
\end{equation}
The boundary conditions (\ref{eq:bcond}) imply two branches of solutions.
One is given by
\begin{equation}
     X={\rm const}~ (1-\cos (kz))~,~~kL=2\pi n~~(n=1,2,...)~,
     \label{eq:bcsol}
\end{equation}
and for the other $k$ has to satisfy $\tan (kL/2)=kL/2$.\\
The critical value $F_{c}$ corresponds to the lowest mode with $n=1$ (no node),
whence
\begin{equation}
     F_{c}=\frac{4\pi ^{2}EI}{L^2}~~(\rm{Euler})~.
     \label{eq:euler}
\end{equation}
For $n=1$ the deflection (\ref{eq:bcsol}) can be written as
\begin{equation}
     X(z)={\rm const} \sin ^{2} \left(\frac{\pi z}{L}\right)~.
     \label{eq:bcsol2}
\end{equation}

When $F$ becomes larger than $F_{c}$, the system, when perturbed 
infinitesimally, jumps to a new ground state in which the U(1) symmetry
is broken (the bent rod). This new state is {\em degenerate}, since
the rod can be bent in any plane containing the $z$ axis. This is a nice 
example of the phenomenon of {\em spontaneous symmetry breaking} (SSB).

The relevance of the criterion (\ref{eq:euler}) for engineering is obvious.
Euler has also studied the shapes of bent rods undergoing large deflections.
 (This is discussed in \cite{Lan}, \S 19.)
\newpage


\section{Galois Theory, the origin of group theory to analyse symmetries}

I come now to an entirely different chapter.

One of the main contributions of Galois was to identify the group concept
and to use it to analyse the problem of solvability of polynomial equations 
by radicals. This enabled him to find a criterion of solvability which has
unsolvability of the general quintic as just one of many corollaries. Galois
theory is in many ways the origin of group theory as a tool to analyse 
symmetries. It may thus not be completely out of place to make a few remarks
about this very beautiful and profound theory, even if it has, so far, no 
{\em direct} applications in physics. (The relations between physics
and pure mathematics are much more subtle than most physicists are aware of.)
Galois Theory nowadays plays an important role in algebraic geometry and 
number theory.

\subsection{Basic concepts and fundamental theorem of Galois Theory}

In Galois Theory one studies {\em field extensions} $K$ of a {\em base field}
$F$. The extension $K$ of $F$ ($F \subset K$) may be regarded as a vector 
space over $F$. We write [$K:F$] for the dimension of $K$ as an $F$-vector 
space and assume always that this is finite. The pair $F \subset K$ is then 
called a {\em finite} extension. The reader may assume (for simplicity) 
that all fields are subfields of the complex numbers $\mathbb{C}$ containing 
the rational numbers $\mathbb{Q}$. A simple example is 
$\mathbb{R} \subset \mathbb{C}$, with $[\mathbb{C} : \mathbb{R}]=2$.\\

Important examples of field extensions arise as follows.
Consider a polynomial $p(X)$ in the indeterminate $X$,
\begin{equation}
     p(X)=a_{0}+a_{1}X+...+a_{n}X^{n},
     \label{eq:polynom}
\end{equation}
whose coefficient are (for instance) in $\mathbb{Q}$. Let 
$\alpha_{1},...,\alpha_{n}$ be the roots of $p(X)$ in $\mathbb{C}$. The 
smallest subfield of $\mathbb{C}$, containing $\mathbb{Q}$ as well as the 
roots $\alpha_{1},...,\alpha_{n}$, is denoted by 
$\mathbb{Q}(\alpha_{1},...,\alpha_{n})$ and is called the {\em splitting field}
of $p$ over $\mathbb{Q}$. (This notion can, of course, be generalised to arbitrary 
fields $F$ and polynomials over $F$, instead of $\mathbb{Q}$.)\\

For finite extensions $F \subset K$ the field $K$ is {\em algebraic}
over $F$, i.e., for every $\alpha \in K$ there is a polynomial $f \in F[X]$,
such that $f(\alpha)=0$.\\

The {\em Galois group}, $Gal(K/F)$, of a field extension $F\subset K$
consists of all automorphisms of $K$ which leave the elements of $F$ fixed.
For finite extensions the Galois group is always finite. Clearly, the fixed
set $Fix(Gal(K/F))$, consisting of all elements of $K$ which are left 
invariant under $Gal(K/F)$, contains $F$, but may in general be larger. We
say that $F \subset K$ is a {\em Galois extension}, if
\begin{equation}
     Fix(Gal(K/F))=F~.
     \label{eq:gal}
\end{equation}
One can show that this is equivalent to
\begin{equation}
     |Gal(K/F)|=[K:F]~.
     \label{eq:gal2}
\end{equation}
(For a finite group $G$, the number of elements is denoted by $|G|$.) This
is just one of several characterizations of Galois extensions.\\

Now we come to a first central result, which provides a key to analyse the 
structure of field extensions with the help of group theory.
\begin{Th}[Fundamental Theorem of Galois Theory.]
     Let $K$ be a finite Galois extension of $F$, and let $G=Gal(K/F)$. Then
there is a 1-1 inclusion reversing correspondence between intermediate fields
$L~(F \subset L \subset K)$ and subgroups of $G$, given by
\begin{equation}
     L \longmapsto Gal(K/L)
\end{equation}
and
\begin{equation}
     H \longmapsto Fix(H)~~(H \subset G)~.
\end{equation}
Furthermore, if $L \longleftrightarrow H$, then $[K:L]=|H|$ and $[L:F]=[G:H]$ 
(= order of $G$ in $H$). Moreover, $H$ is a normal subgroup of $G$ if and 
only if $L$ is
Galois over $F$. When this occurs,
\begin{equation}
     Gal(L/F) \cong G/H~.
\end{equation}
\end{Th}

\subsection{Solutions by radicals}
Consider the polynomial equation $p(X)=0$, with a polynomial $p(X)$ of the
form (\ref{eq:polynom}). The elements 
obtained from
$a_{0},...,a_{n}$ by the operations $+,\cdot,\div$ form the 
{\em coefficient field} $\mathbb{Q}(a_{0},...,a_{n})$. An element obtained
from this field by a finite number of roots 
$ ~\sqrt[\leftroot{2}\uproot{4}  ]{~~},\sqrt[\leftroot{2}\uproot{4} 3]{~~},
\sqrt[\leftroot{2}\uproot{4} 4]{~~},...$ lies in an extension 
field of $\mathbb{Q}(a_{0},...,a_{n})$ obtained by a finite number of 
{\em radical adjunctions}. We say that adjunctions of an element 
$\alpha$ to a field $F$ is {\em radical} if there is a positive integer 
$m$ such that $\alpha^{m} \in F$. The result of several radical adjunctions
$F(\alpha_{1})...(\alpha_{k})$ is called a {\em radical extension} of
$F$ and is denoted by $F(\alpha_{1},...,\alpha_{k})$.

Thus the problem of {\em solution by radicals} is to find a radical 
extension of the coefficient field $\mathbb{Q}(\alpha_{0},...,\alpha_{n})$ which 
includes the roots $x_{1},...,x_{n}$ of
\begin{equation}
     a_{0}+a_{1}x+...+a_{n}x^{n}=0.
\end{equation} 
For example, the formula for the solution of the quadratic equation with
$a_{2}=1$,
\begin{equation}
     x_{1,2}=\frac{-a_{1}\pm \sqrt{a_{1}^{2}-4a_{0}}}{2}~, \nonumber
\end{equation}
shows that $\mathbb{Q}(x_{1},x_{2})$ is contained in the radical extension 
$\mathbb{Q}(a_{0},a_{1},\sqrt{a_{1}^{2}-4a_{0}})$.\\

A classical application of Galois Theory, and one of the main results of
Galois himself, is the following
\begin{Th}[Galois]
A polynomial $f \in \mathbb{Q}[X]$ with splitting field $K$ over $\mathbb{Q}$ is solvable 
by radicals if and only if the Galois group $Gal(K/\mathbb{Q})$ is solvable. (Here,
$\mathbb{Q}$ can be replaced by any field of characteristic 0.) 
\end{Th}

We recall that a group $G$ is {\em solvable} if there is a chain of 
subgroups
\begin{equation}
     <e>=H_{0}\subseteq H_{1}\subseteq ...\subseteq H_{n}=G~, \nonumber
\end{equation}
such that for all $i$, the subgroup $H_{i}$ is normal in $H_{i+1}$ and the
quotient group $H_{i+1}/H_{i}$ is Abelian.

Consider, for instance, the equation
\begin{equation}
     x^{5}-4x+2=0~.
     \label{eq:galoisex}
\end{equation}
It is not difficult to show that the Galois group of this equation (i.e., the 
Galois group $Gal(K/\mathbb{Q})$, $K$ being the splitting field of $f(X)= X^{5}-4X+2$)
is the permutation group $S_{5}$, which is not solvable. (The latter fact can
be proven in a very elementary way.) Galois' theorem thus implies that
(\ref{eq:galoisex}) is {\em not solvable by radicals}.

\subsection{Ruler and compass constructions}

Galois Theory can also be used to answer some ancient questions concerning
constructions with ruler and compass.
Examples are:
\begin{itemize}
    \item[(i)]{Is it possible to trisect any angle?}
    \item[(ii)]{Is it possible to double the cube?}
    \item[(iii)]{For which $n$ is it possible to construct a regular $n$-gon?}
\end{itemize}

Let me only address the third question, whose solution makes use of much of 
Galois Theory. Consider first the case when $n$ is a prime number $p$. In
this case we have the
\begin{Th}
      A regular $p$-gon ($p$ a prime number) is constructible if and only if
$p-1$ is a power of 2.
\end{Th}

Such numbers are called {\em Fermat primes}. Unfortunately, we do not
know whether there are any Fermat primes beyond
\begin{equation}
     3,5,17,257,65537~.
\end{equation}
Deciding whether 65537 is the last Fermat prime may well tax the best
mathematician of the future. Fermat had conjectured (1640) that
$2^{2^{h}}+1$ is prime for all natural numbers $h$, but Euler found (1738)
that 641 divides $2^{2^{5}}+1$.\\

The theorem above implies, for example, that a regular 17-gon is constructible.
An explicit construction was given by the 19-year old Gauss in 1796.\\

For the general case we have to introduce the {\em Euler phi function}
$\phi(n)$. This counts those integers among $1,2,...,n-1$ which have no 
nontrivial common divisors with $n$. (For a prime, we clearly have 
$\phi(p)=p-1$.) The last theorem generalizes to:
\begin{Th}
     A regular $n$-gon is constructible if and only if $\phi(n)$ is a power
of 2.
\end{Th}

If $n=p_{1}^{m_{1}}...p_{r}^{m_{r}}$ is the prime factorization of $n$, then
\mbox{$\phi(n)=\prod_{i} p^{m_{i}-1} (p_{i}-1)$.} It is not difficult to 
show that $\phi(n)$ is a power of 2 if and only if
\begin{equation}
     n=2^{s} q_{1}...q_{r}~,
\end{equation}
where the $q_{i}$ are Fermat primes. In this sense problem (iii) is solved.
But remember, we may not know all Fermat primes.

\subsubsection*{Discussion}

Galois' work was to sketchy to be understood by his referees, and it was not
published in his brief lifetime (he died after a duel in 1832, aged 20). It
was later published by Liouville in 1846, after he became convinced that
Galois' proof of his solvability criterion of equations was correct. Over the
next two decades the group concept was then assimilated to the point where
Jordan could write his "Trait$\acute{{\rm e}}$ des substitutions et des 
$\acute{{\rm e}}$quations alg$\acute{{\rm e}}$brique"
in 1870. This book is inspired by Galois Theory, but group theory takes over 
almost completely.\\

In the 1870s geometry also began to influence group theory. Of great 
importance was Klein's {\em Erlanger Programm} in 1872, which emphasized
the unifying role of groups in geometry. In those days nobody could imagine
that group theory would one day play also in physics such a decisive and
increasingly important role.

Among the many excellent textbooks on Galois Theory, I refer to the recent one
by J. Stillwell \cite{Sti94}, which is written in a lively style, emphasizing
the historical context, and avoiding unnecessary generalizations (for
beginners).


\section{Rotating selfgravitating equilibrium figures}

For nearly a century it was believed that Maclaurin's axially symmetric
ellipsoids (1742) represent the only admissable solutions of the problem of the
equilibrium of selfgravitating uniformly rotating homogeneous masses.\\

In 1834 Jacobi came up with the surprising announcement:\\

{\em "One would make a grave mistake if one supposed that the spheroids of
revolution are the only admissable figures of equilibrium even under the 
restrictive assumption of second degree surfaces. $\cdot \cdot \cdot$ In fact 
a simple consideration shows that ellipsoids with three unequal axes can 
very well be figures of equilibrium; and that one can assume an ellipse of 
arbitrary shape for the equatorial section and determine the third axes (which
is also the least of the three axes) and the angular velocity of rotation 
such that the ellipsoid is a figure of equilibrium."}\\

Jacobi's surprising discovery can be regarded as an example of spontaneous 
symmetry breaking of the group U(1). For small angular momenta there is only 
the symmetric solution, but above some bifurcation point there exists also 
an unsymmetric solution. Before saying more about this and the stability 
issue, let me enter a bit into the preceding history, which begins with 
Newton.\\

In the Principia, Book III, Propositions XVIII-XX, Newton derives the 
oblateness of the Earth and other planets. I first give his result.

Let
\begin{equation}
     \epsilon=\frac{\rm{equatorial~radius-polar~radius}}{\rm{mean~radius~(R)}}
\end{equation} 
be the ellipticity, $M$ the total mass, $\Omega $ the angular velocity, and
$R$ the average radius of the planet. With a beautiful argument (described
below) Newton finds that
\begin{equation}
     \epsilon=\frac{5}{4}\frac{\Omega^{2}R^{3}}{GM}~,
\end{equation}
if the body is assumed to be homogeneous.\\

In his derivation, Newton imagines a hole of unit cross-section drilled from 
a point of the equator to the center of the Earth and a similar hole from 
the pole to the center. Both 'canals' are imagined to be filled with water
(see Fig.2).
Newton studies now the implication of the equilibrium condition:
\begin{equation}
   \rm{weight~of~equilibrium~column=weight~of~polar~column}.
   \label{eq:eqcond}
\end{equation}
\begin{figure}[H]
\begin{center}
\includegraphics[height=0.25\textheight, angle=0]{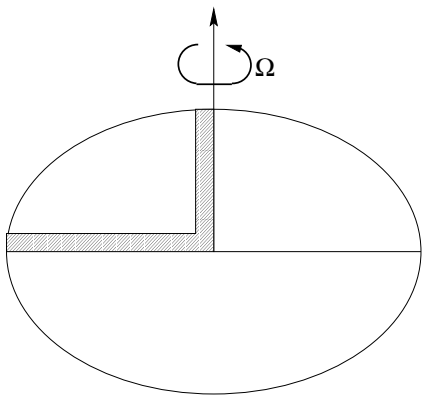}
\caption{\small{Newton's water 'canals' through the Earth.}}
\end{center}
\label{fig:canal}
\end{figure}
\noindent
Along the equator the acceleration due to gravity is 'diluted' by the 
centrifugal acceleration. Newton has shown earlier (Book I, Propositions
LXXIII and Corollary III, Proposition XCI) that both these accelerations,
in a homogeneous body, vary from the center linearly with the distance. 
Therefore, the dilution factor remains constant and is equal to its value, 
$m$, 
on the boundary
\begin{equation}
     m=\frac{\rm{centrifugal~acceleration~at~the~equator}}
     {\rm{mean~gravitational~acceleration~on~the~surface}}=
     \frac{\Omega^{2}R}{GM/R^{2}}~.
\end{equation}
(Here, $m<<1$ is used.)\\

Now, the weight of the equatorial column is equal to
$\frac{1}{2}\rho a g_{\rm{eq}}(1-m)$ ($a$=equatorial radius), and the weight
of the polar column is $\frac{1}{2}\rho c g_{\rm{pole}}$ ($c$=polar radius).
Thus eq.(\ref{eq:eqcond}) gives
\begin{equation}
     a g_{\rm{eq}}(1-m)=c g_{\rm{pole}}~.
\end{equation}
Newton recognizes that this equation is valid for any $\epsilon$!\\
Using also $c=(1-\epsilon)a$ this gives
\begin{equation}
     \frac{g_{\rm{pole}}}{g_{\rm{eq}}}=\frac{a}{c}(1-m)=\frac{1-m}{1-\epsilon}
     \simeq (1+\epsilon-m)~.
     \label{eq:poleq}
\end{equation}

Chandrasekhar explains in his beautiful last book (\cite{Cha95}, p. 386), how
Newton arrived at the relations
\begin{equation}
     g_{\rm{eq}}=\frac{4\pi}{3}a(1-\frac{2}{5}\epsilon),~~
     g_{\rm{pole}}=\frac{4\pi}{3}a(1-\frac{1}{5}\epsilon)~.
\end{equation}
(For us it is easier to derive this from the Maclaurin solutions.)\\
If this is used in (\ref{eq:poleq}), Newton's result
\begin{equation}
     \epsilon=\frac{5}{4}m
\end{equation}
is obtained.

It was known already in Newton's time that for the Earth $m\simeq 1/290$.
Therefore, Newton concludes that if the Earth were homogeneous, it should be
oblate with an ellipticity
\begin{equation}
     \epsilon\simeq\frac{5}{4}\frac{1}{290}\simeq\frac{1}{230}~.
\end{equation}
We know that the actual ellipticity of the Earth is substantially smaller
($\simeq$1/294). This is interpreted in terms of the inhomogeneity of the
Earth.\\

\begin{figure}[h]
\begin{center}
\includegraphics[height=0.3\textheight, angle=0]{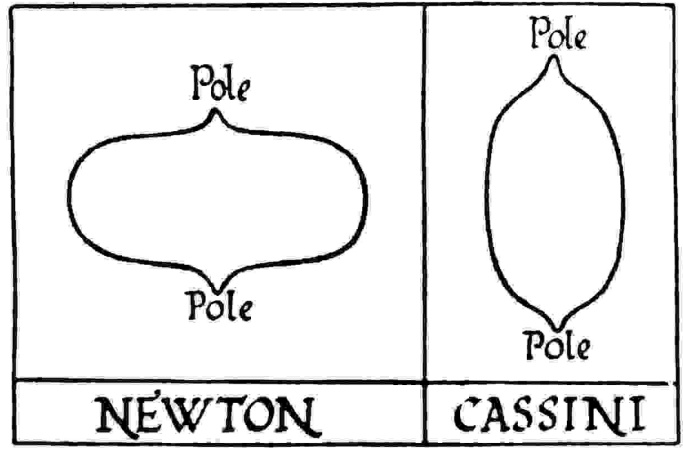}
\caption{\small{Caricature of the controversy on the figure of the Earth.}}
\end{center}
\label{fig:cassini}
\end{figure}

Newton's prediction was against the ideas of the Cassini school, as is 
illustrated in the old-time caricature below
(Fig.\ref{fig:cassini}). 
The famous controversy was finally settled by a measurement in 1738 of the
arc of the meridian by a French expedition to Lapland, led by Maupertuis.
This was an exceedingly difficult and tedious enterprise.\\

I. Todhunter writes in his "A history of the mathematical theories of
attraction and the figures of the Earth" in 1873 (reprinted by Dover 
Publications, p. 100):\\

{\em "The success of the Arctic expedition may be fairly ascribed in great 
measure 
to the skill and energy of Maupertuis: and his fame was widely celebrated.
The engravings of the period represent him in the costume of a Lapland
Hercules, having a fur cap over his eyes; with one hand he holds a club, and
with the other he compresses a terrestrial globe. Voltaire, then his friend,
congratulated him warmly for having 'aplati les p\^oles et les Cassini'."}\\
\newline
Maupertuis' report to the Paris Academy became a bestseller. A German
translation appeared 1741 in Z\"urich.\\

I have to refrain to tell you more about Maupertuis, who later became the first
President of the Prussian Academy, founded by Frederick the Great. Maupertuis
was actually an organizer, but not a great scientist. His end was tragic.\\

After this long digression I come back to Jacobi's solution. 
In Fig.4
the moment of inertia, $\Theta$, relative to the rotation axis (in
units of the non-rotational case) is shown as a function of the angular 
velocity squared (in units of $\pi G \rho$, $\rho$=uniform density) for the
Maclaurin and the triaxial Jacoby solutions. The latter sequence bifurcates
from the axially symmetric family at the point where 
$\Omega^{2}/\pi G \rho=0.37423$ (eccentricity $\epsilon $=0.81267). One sees
from Fig.4 that for $\Omega^{2}/\pi G \rho<0.37423$ there are 
three 
equilibrium figures possible: two Maclaurin spheroids and one Jacobi ellipsoid;
for $0.4493>\Omega^{2}/\pi G \rho>0.3742$ only the Maclaurin figures are 
possible; and finally for $\Omega^{2}/\pi G \rho>0.4493$ there are no 
equilibrium solutions. (This enumeration was given by C.O. Meyer in 1842.)\\

As Riemann has shown, the Maclaurin ellipsoids become unstable in point $B$
in Fig.4, where $\Omega ^{2}/\pi G \rho=0.4402$. 
Poincar$\acute{\rm{e}}$
and Cartan proved that the Jacobi sequence becomes unstable at 
$\Omega^{2}/\pi G \rho=0.2840$.\\

\begin{figure}[H]
\begin{center}
\includegraphics[height=0.5\textheight, angle=0]{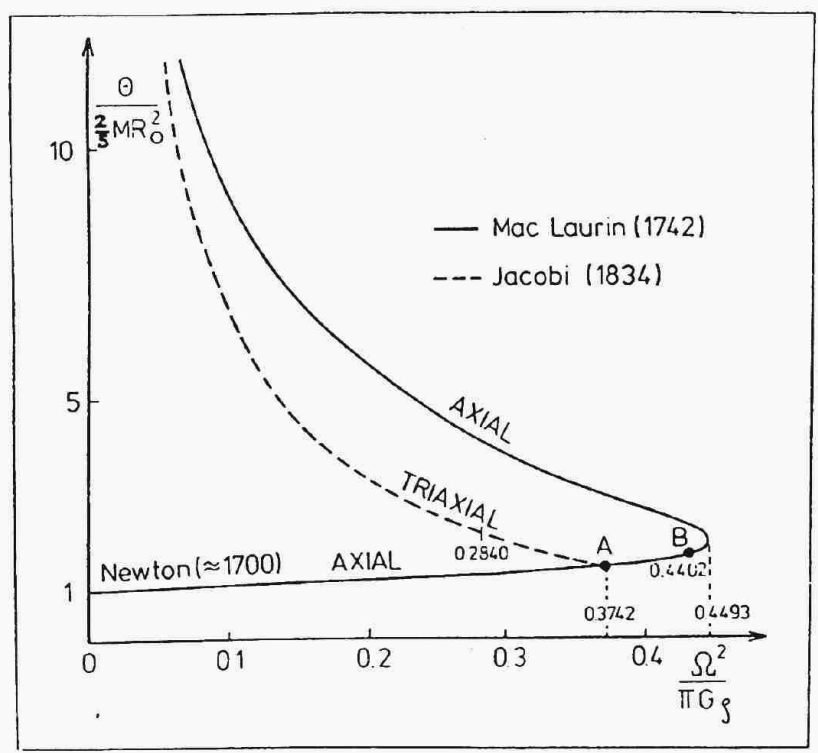}
\caption{\small{Maclaurin and Jacobi sequences of ellipsoidal equilibrium 
figures.}}
\end{center}
\label{fig:jacobi}
\end{figure}

S. Chandrasekhar has devoted an entire book on the ellipsoidal figures of 
equilibrium and their stability analysis \cite{Cha87}. In Chapter 1 he gives 
a detailed discussion of the interesting history of this subject, to which 
an impressive list of great mathematicians, from Newton to Cartan, has 
contributed over a long period of time.


\section{Spontaneous symmetry breaking due to thermal instabilities}

Thermal instabilities often arise when a fluid is heated from below. A
classical example is a horizontal layer of fluid with its lower side hotter
than its upper. Due to thermal expansion, the fluid at the bottom will be 
lighter than at the top. When the temperature difference across the layer is 
great enough the stabilizing effects of viscosity and thermal conductivity
are overcome by the destabilizing buoyancy, and an overturning instability
ensues as thermal convection.\\

Such a convective instability seems to have been first described by James
Thomson (1882), the elder brother of Lord Kelvin, but the first quantitative
experiments were made by B$\acute{\rm{e}}$nard (1900). Stimulated by these 
experiments, Rayleigh formulated in 1916 the theory of convective instability 
of a layer of fluid between horizontal planes. \\

Starting from the basic hydrodynamic equations (in the Boussinesq 
approximation) and the boundary conditions, Rayleigh derived the linear 
equations for normal modes about the equilibrium solution. He then showed 
that an instability sets in when the following dimensionless parameter
\begin{equation}
     R=g \alpha \beta d^{4}/\kappa \nu
\end{equation}
exceeds a certain critical value $R_{c}$. Here $g$ is the acceleration due 
to gravity, $\alpha$ the coefficient of thermal expansion of the fluid, 
$\beta$ the magnitude of the vertical temperature gradient of the basic 
state at rest, $d$ the depth of the layer of the fluid, $\kappa$ the thermal 
conductivity and $\nu$ the kinematic viscosity. The parameter $R$ is now 
called the {\em Rayleigh number}.\\

If both boundaries are rigid, the critical value turns out to be
\begin{equation}
     R_{c}=1707.762~.
\end{equation}
(A detailed derivation can be found in \cite{Cha81}, Chap.II. This
beautiful book gives also the relevant references to the original literature.)
Experimentally one found
\begin{equation}
     R_{c}^{\rm{exp.}}=1700\pm 51~,
\end{equation}
in complete agreement with the theoretical value.\\

There is an ironic twist to what is cold 
{\em B$\acute{\rm{e}}$nard convection}. Most of the motions observed by
B$\acute{\rm{e}}$nard, being in very thin layers with a free surface, were 
actually driven by variations of the surface tension with temperature and
{\em not} by a thermal instability of a light fluid below a heavy one.
This effect of the surface tension becomes, however, unimportant if the 
thickness of the layer is sufficiently large.\\
\begin{figure}[h]
\begin{center}
\includegraphics[height=0.4\textheight, angle=-90]{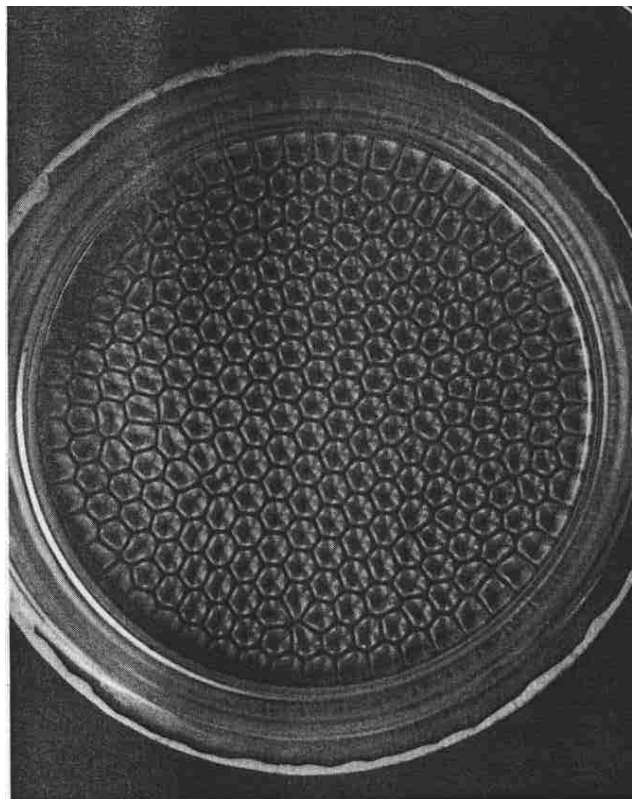}
\caption{\small{B$\acute{\rm{e}}$nard cells under an air surface.}}
\end{center}
\label{fig:benard}
\end{figure}

When $R$ becomes larger then $R_{c}$, the motion of the fluid assumes a
stationary, cellular character (spontaneous breakdown of translational 
symmetry). If the experiment is performed with sufficient care, the cells 
become equal and often form a regular hexagonal pattern 
(see Fig.5). As the 
Rayleigh number increases, a series of transitions from one complicated flow
to the next more complicated one can be detected. An understanding of all this
is difficult, because nonlinearities become significant.\\

This concludes my sundry of ancient examples on SSB.


\section{Goldstone- and Mermin-Wagner theorems}

Before Onsager had found his famous exact solution of the 2-dimensional Ising 
model, it was not generally accepted that the rules of statistical mechanics 
are able to describe phase transitions. As late as 1937, at the Van der Waals 
Centenary Conference, there was lively debate on whether phase transitions 
are consistent with the formalism of statistical mechanics. After the debate, 
Kramers suggested that a vote should be taken, on whether the infinite-volume 
limit could provide the answer. The result of that vote was close, but the 
infinite-volume limit did finally win. (More on this can be found in Pais' 
wonderful biography of Einstein \cite{Pai82}, pp. 432-33.)

We now know that first order phase-transitions in some parameter are equivalent
to the existence of more than one translational invariant infinite-volume
equilibrium state. This subject has matured very much, especially by the many 
advances in the sixties and seventies.

Lattice approximations of Euclidean formulations of quantum field theories are 
{\em classical} statistical mechanics systems. The simplest example, a 
(multicomponent) scalar field theory, leads to a spin model with 
ferromagnetic nearest neighbor coupling. This alone is good reason for 
studying lattice spin models.

\subsection{Spontaneous symmetry breaking for the Ising model (d$\geq$2)}

The Ising model illustrates very nicely the phenomenon of SSB. I recall that 
the configurations of this model consist of distributions of spins 
${\sigma_{x}=\pm 1}$ at each lattice point $x$ of a hypercubic lattice 
$\mathbb{Z}^{d}$, say. The interaction is invariant under the group 
$\mathbb{Z}_{2}$, consisting of the identity and the reflection 
$\sigma_{x}\rightarrow -\sigma_{x}$ for all lattice sites $x$.\\

Above a critical temperature $T_{c}$ there is only {\em one} 
infinite-volume equilibrium state (state=probability measure). However, for 
$T<T_{c}$ each translationally invariant equilibrium state $\mu^{\beta}$ 
($\beta=1/kT$) is a convex linear combination of {\em two different} 
extremal states $\mu_{\pm}^{\beta}$:
\begin{equation}
     \mu^{\beta}=\lambda \mu_{+}^{\beta}+(1-\lambda)\mu_{-}^{\beta}
     \qquad (0\leq\lambda \leq 1)~.
\end{equation}
This means that $\mu^{\beta}$ describes a {\em mixture} of two 
{\em pure phases}. The latter probability measures $\mu_{\pm}^{\beta}$
are weak limits of Gibbs states on finite regions 
$\Lambda\subset \mathbb{Z}^{d}$ with $\pm$ boundary conditions outside 
$\Lambda$. Since they are different, they are not invariant under the 
symmetry group $\mathbb{Z}_{2}$ of the interaction; the symmetry is 
{\em spontaneously broken for these pure phases}. Correspondingly, 
the spontaneous magnetizations
\begin{equation}
     m_{\pm}(\beta)=<\sigma_{x}>_{\mu_{\pm}^{\beta}}
\end{equation}
do not vanish for $\beta>\beta_{c}$ (see Fig.6).
\begin{figure}[h]
\begin{center}
\includegraphics[height=0.3\textheight, angle=0]{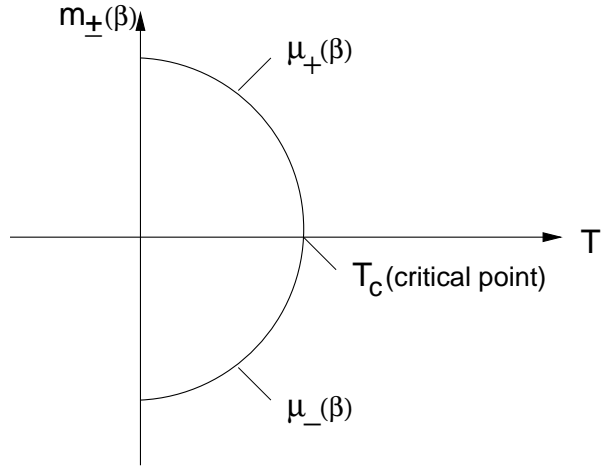}
\caption{\small{Spontaneous magnetization in the Ising model for $d\geq 2$.}}
\end{center}
\label{fig:sponmagn}
\end{figure}

\subsection{Spin systems with continuous symmetry groups}

Instead of the discrete spin variables $\sigma_{x}=\pm 1$, we consider now 
continuous 'spins': $x\mapsto \varphi_{x}\in \mathbb{R}^{N},S^{N},...,$ and 
interactions which are invariant under a continuous symmetry group.\\

Let $\mu$ be a translationally invariant infinite-volume equilibrium state. 
Assume that the following {\em cluster property} holds for local 
observables $A,~B$ (that is, observables which depend only on finitely many 
spin variables)
\begin{equation}
     \arrowvert <A\tau_{x}(B)>-<A><B>\arrowvert=
     {\cal O} \left( \frac{1}{|x|^{\delta}}\right)~,\qquad \delta > 0~,
\end{equation}
where $\tau_{a}$ denotes the translation by $a$. One can prove that this 
implies the following, provided the interaction has finite range (this can 
be weakened):
The equilibrium state $\mu$ is {\em invariant} under the symmetry
 group, if
\begin{equation}
     \delta>d-2~.
\end{equation}
For a proof, see, e.g., Ref. \cite{Mar82}, and references therein.
\subsubsection*{Consequences}

\begin{itemize}
    \item[1.]{If {\em $d=2$}  {\em all clustering states are 
invariant}, i.e., the continuous symmetry group $G$ cannot be spontaneously 
broken ({\em Mermin-Wagner Theorem}).}
    \item[2.]{Consider the case {\em $d=3$}. Then in any nonsymmetric 
phase the clustering cannot decay faster than $|x|^{-1}$.}
    \item[3.]{For {\em $d=4$} (field theory) there is 
{\em no mass gap in a nonsymmetric phase} (otherwise there would be an 
exponential clustering, which is not possible). This is the 
{\em Goldstone-Theorem} (for lattice models).}
\end{itemize}


\section{Order parameters and Elitzur's theorem for gauge theories}

In the previous section we considered systems with global symmetries. A 
spontaneous breaking of such a symmetry is accompanied by a nonvanishing 
spontaneous magnetization. At first sight one expects something similar for 
gauge theories. However, Elitzur has shown that {\em local} quantities, 
like a Higgs field, which are not gauge invariant, have always vanishing mean 
values; that is, {\em local} observables {\em cannot} exhibit 
spontaneous breaking of {\em local} gauge symmetries.\\

Since this is quite easy to prove, I give here the details for lattice gauge 
models. It is instructive to consider a lattice gauge theory with the gauge 
group $\mathbb{Z}_{2}$, because this shows the contrast to the Ising model.\\

First, I need some standard notation. A field configuration is a map of bonds 
($b$) into the gauge group $\mathbb{Z}_{2}$: $b\mapsto \sigma_{b}(=\pm 1)$. 
$\sigma_{\partial P}$ denotes the group element for a plaquette $P$. The 
action for a finite region $\Lambda$ of the lattice $\mathbb{Z}^{d}$ is
\begin{equation}
     S_{\Lambda}(\{ \sigma \})=-\sum_{P\subset \Lambda}\sigma_{\partial P}-
     h\sum_{b\subset \Lambda}\sigma_{b}~,
\end{equation}
where $h$ is an external 'field'. The expectation value for a local 
observable \vspace{5mm}$A$\vspace{-5mm} is
\begin{equation}
     <A> _{\Lambda} =Z_{\Lambda}^{-1}\sum_{\{ \sigma \}}A(\{ \sigma \})
     e^{-\beta S_{\Lambda}(\{ \sigma \})}~,
\end{equation}
where $Z_{\Lambda}$ is the partition sum
\begin{equation}
     Z_{\Lambda}=\sum_{\{ \sigma \}} e^{-\beta S_{\Lambda}(\{ \sigma \})}~.
\end{equation}
In sharp contrast to what happens in the Ising model, the mean value of 
$\sigma_{b}$ does not signal a symmetry breaking:
\begin{Th}[Elitzur]
     For the expectation value $<\sigma_{b}>_{\Lambda}(h)$ we have
\begin{equation}
     \lim_{h\downarrow 0}<\sigma_{b}>_{\Lambda}(h)=0~~\rm{uniformly~in~}
     \Lambda ~\rm{and}~\beta~.
     \label{eq:limexpval}
\end{equation}
In particular, the thermodynamic limit of $<\sigma_{b}>_{\Lambda}$ vanishes 
for $h\downarrow 0$ (no spontaneous 'magnetization').
\end{Th}

Before giving the simple proof, I remark that it is easy to add a Higgs field 
$\phi_{x}$ to the model, and prove similarly that
\begin{equation}
     <\phi_{x}>=0~.
     \label{eq:higgs}
\end{equation}
This does, however, not exclude a Higgs phase with exponential fall off of 
the correlation function $<\sigma_{\partial P_{1}}\sigma_{\partial P_{2}}>$ 
for the gauge fields (mass generation), but (\ref{eq:limexpval}) and 
(\ref{eq:higgs}) show that this is not signaled by local observables.

\subsubsection*{Proof of Elizur's theorem.}

We choose in (\ref{eq:limexpval}) the bond variable $\sigma_{01}$ for the 
bond $b=<0,1>$ and estimate in
\begin{equation}
     <\sigma_{01}>_{\Lambda}=\frac{\sum_{\{ \sigma \}}\sigma_{01}
     \exp \left(\beta \sum_{P}\sigma_{\partial P}+\beta h \sum_{b}
     \sigma_{b}\right)}{\sum_{\{ \sigma \}} \exp\left(\beta \sum_{P}
     \sigma_{\partial P}+\beta h \sum_{b}\sigma_{b}\right)}
\end{equation}
the numerator $N$ and the denominator $D$ separately.\\

Consider a gauge transformation 
$\sigma_{ij}'=\epsilon_{i}\sigma_{ij}\epsilon_{j}$, with $\epsilon_{i}=1$ for 
$i\neq 0$ and replace $\sigma_{ij}$ in $N$ and $D$ by 
$\epsilon_{i}\sigma_{ij}'\epsilon_{j}$, dropping afterwards the prime of 
$\sigma_{ij}$ ($\sigma_{ij}\rightarrow \epsilon_{i}\sigma_{ij}\epsilon_{j}$). 
This can be done for $\epsilon_{0}=\pm 1$. $D$ is equal to half of the sum:
\begin{equation}
     D=\frac{1}{2}\sum_{\epsilon_{0}=\pm 1}\sum_{\{ \sigma \}}
     \exp\left(\beta \sum_{P}\sigma_{\partial P}+\beta h \sideset{}{'}\sum_{b}
     \sigma_{b}
     \right)\exp\left(\epsilon_{0}\beta h \sum_{j=1}^{2d}\sigma_{0j}\right)~,
\end{equation}
where the prime of the sum means that the bonds $<0,j>$ must be excluded. 
Clearly, for $h>0$
\begin{equation}
     D\geq \sum_{\{ \sigma \}}\exp\left(\beta \sum_{P}\sigma_{\partial P}
     +\beta h \sideset{}{'}\sum_{b} \sigma_{b}\right) ~ \frac{1}{2}~
     e^{-2d\beta h}~.
\end{equation}
Similarly, we have for the numerator
\begin{equation}
     |N|\leq \sum_{\{ \sigma \}}\exp\left(\beta \sum_{P}\sigma_{\partial P}
     +\beta h 
     \sideset{}{'}\sum_{b} \sigma_{b}\right) ~ \frac{1}{2} ~\left| 
     \sum_{\epsilon_{0}=
     \pm 1}\epsilon_{0}\sigma_{01}\exp\left(\epsilon_{0}\beta h \sum_{j=1}^{2d}
     \sigma_{0j}\right) \right| ~,
\end{equation}
and thus
\begin{equation}
     |N|\leq \sum_{\{ \sigma \}}\exp\left(\beta \sum_{P}\sigma_{\partial P}
     +\beta h 
     \sideset{}{'}\sum_{b}\sigma_{b}\right)\sinh(2d\beta h)~.
\end{equation}
This gives the estimate
\begin{equation}
     \arrowvert <\sigma_{01}>_{\Lambda}(h)\arrowvert \leq 2 e^{2d\beta h}
     \sinh(2d\beta h) \overset{\small{(h \downarrow 0)}}{\longrightarrow} 0~,
\end{equation}
\hspace{8cm} {\em uniformly} in $\Lambda$ and $\beta$. \hspace{0.5cm}  QED.
\vspace{0.5cm}

This argument can easily be generalized to any lattice gauge theory and any 
local observable which is noninvariant under the gauge group 
(see, e.g., \cite{Itz89}, Chap. 6).\\

Elitzur's theorem implies that possible order parameters have to be nonlocal 
objects. An example of such a quantity is the Wilson loop and the associated 
string tension.\\

What is the reason for this different behavior of models with local and 
global symmetries? Consider again the Ising model in the absence of an 
external field. At low temperatures, the two regions in configuration space 
with opposite magnetizations ${\sigma_{x}}$ and ${-\sigma_{x}}$ can only be 
connected by a dynamical path involving the creation of an infinite interface 
which costs an infinite amount of energy. Therefor the process cannot occur 
spontaneously. Alternatively, one can make use of a small external field 
which is switched off {\em after} the thermodynamic limit is taken. On 
the other hand, in a local gauge theory one can perform gauge transformations 
which act only on a finite set of basic variables on which a local 
"observable" 
depends, and which leaves the complementary set invariant.

\end{document}